\documentclass[aps,prb,twocolumn,groupeaddress]{revtex4}

\usepackage{graphicx}
\usepackage{dcolumn}

\begin{document}


\title{
Parity effect in superconducting aluminum single electron transistors 
with spatial gap profile controlled by film thickness
}


\author{T. Yamamoto, Y. Nakamura}
\author{Yu.\ A. Pashkin}
 \altaffiliation{on leave from Lebedev Physical Institute, Moscow 119991, Russia}
\author{O. Astafiev and J. S. Tsai}
\affiliation{
NEC Fundamental and Environmental Research Laboratories, 34 Miyukigaoka, Tsukuba, 
Ibaraki 305-8501, Japan\\
and Frontier Research System, RIKEN, 2-1 Hirosawa, Wako, 
Saitama 351-0198, Japan
}


\date{February 2, 2006}

\begin{abstract}
We propose a novel method for suppression of quasiparticle poisoning in Al Coulomb blockade devices. 
The method is based on creation of a proper energy gap profile along the device. 
In contrast to the previously used techniques, the energy gap is controlled by the film thickness. 
Our transport measurements confirm that the quasiparticle poisoning is suppressed and 
clear 2$e$ periodicity is observed only when the island is made much thinner than the leads. 
This result is consistent with the existing model and provides a simple 
method to suppress quasiparticle poisoning. 
\end{abstract}

\pacs{}

\maketitle

\begin{figure}
\includegraphics[width=0.9\columnwidth,clip]{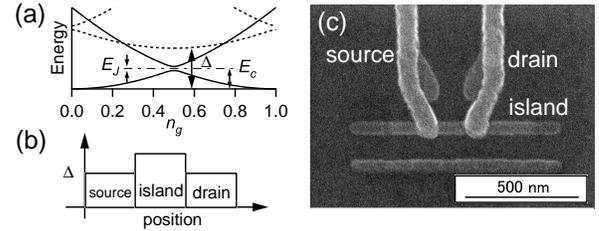}
\caption{\label{fig:fig1}
(a) Schematic diagram of the energy bands of the charge qubit as a function of $n_g$. 
Dashed lines represent the energy bands for the odd parity state. 
(b) Spatial superconducting gap profile, which is favorable for 
the QP-poisoning-free SET device. 
(c) Scanning electron micrograph of the SET sample, 
whose island and leads are 10 and 50 nm thick, respectively. 
The gate electrode is not seen in this picture. 
}
\end{figure}

Parity effect in a small superconducting island, whether it contains an unpaired 
electron (quasiparticle, QP) or not, has been extensively studied in 1990s.~\cite{Averin92,Tuominen92,Lafarge93,Schon94,Joyez94} 
Recently, this topic has attracted research interest again~\cite{Aumentado04,Mannik04,Gunnarsson04,Naaman06} 
because of its importance for the quantum bit (qubit) application of the small Josephson 
junction circuits.~\cite{Nakamura99,Makhlin01,Vion03} 
It is well known that the superconducting charge qubit is least sensitive to the 
main decoherence source, i.e., background charge fluctuation, at the charge degeneracy point~\cite{Vion03}  
where the electrostatic energy of two charge states differing by 2$e$ are equal ($e$: single electron charge). 
This is because the first derivative of the excitation energy of the qubit with respect to the reduced gate charge $n_g=C_gV_g/2e$
becomes zero at this point $n_g=0.5$. 
It should be noted that the curvature of the energy band, which can be utilized for the readout~\cite{Sillanpaa05,Duty05} 
or the coupling~\cite{Averin03} of the qubits, is maximum also at this point. 
Thus, it is important to operate the qubit at the degeneracy point. 
As long as the equilibrium states are concerned, operation at the degeneracy point can be achieved 
if $E_c-E_J/2$ is smaller than $\Delta$, where $E_c$, $E_J$, and $\Delta$ are the single electron charging energy, 
the Josephson energy of a single junction, and the superconducting energy gap, respectively (see Fig.~\ref{fig:fig1}a). 
In this case, the ground state is always even-parity (no QP) state and 
its energy is perfectly 2$e$ periodic in $n_g$ (maximum at half odd integer), which 
should produce 2$e$ periodicity of, for example, the gate modulation of the supercurrent in the 
superconducting single electron transistor (SET). 
In reality, however, it is not a trivial task at all to observe this 2$e$ periodicity even for the 
island having $E_c-E_J/2 < \Delta$ probably due to the existence of nonequilibrium QPs.~\cite{Aumentado04} 


Recently, Aumentado {\it et.\ al.} reported~\cite{Aumentado04} that it is possible to reproducibly 
obtain 2$e$ periodicity of the supercurrent in the superconducting Al SET 
by making $\Delta$ of the island higher than that of the leads (Fig.~\ref{fig:fig1}b), 
because it works as a barrier, which prevent nonequilibrium QPs in the leads from entering the island. 
They realized such a spatial profile of $\Delta$ by introducing a small amount of oxygen during the evaporation of the island. 
Similar effect has been reported for a Cooper-pair box,~\cite{Gunnarsson04} where the required spatial profile is 
obtained by a weak magnetic field applied parallel to the film. 
Motivated by these works, we investigate an alternative way of creating a proper $\Delta$ profile. Our approach is based on the 
fact that $\Delta$ of an Al thin film depends also on the film thickness.~\cite{Townsend72} 
This suggests that we can obtain the spatial profile of $\Delta$ by choosing a proper thickness for the island and the leads. 
This technique is simpler than controlling oxygen pressure or magnetic field. Moreover, improper choice of the thickness may lead to 
the QP-poisoned devices. Our results show that the $\Delta$ profile controlled by the film thickness can change the 
degree of QP poisoning. 

The SET samples were fabricated by electron-beam lithography and double-angle evaporation of Al 
using a trilayer resist (methylmethacrylate-methacrylic acid (MMA-MAA)/Ge/poly-methyl-methacrylate (PMMA), 
200/20/50 nm thick) on an oxidized Si wafer with 
pre-deposited gold pads. The pressure inside the chamber before Al evaporation does not exceed 
$4\times10^{-8}$ Torr, and increases up to $1\times10^{-7}$ Torr during the evaporation. 
No oxygen doping during the Al evaporation was performed in this study. 
The surface of the first Al layer was oxidized by 70 mTorr oxygen for 0.7 to 5 min. 
Figure~\ref{fig:fig1}c shows a scanning electron micrograph of one of the SET devices with the 
thickness of 10 nm and 50 nm for the island and the leads, respectively. The bright area is the evaporated Al. 
The source, drain and gate electrodes are connected to the gold pads, which are about 20 ${\mu}$m away. 
No QP traps~\cite{Joyez94} are used in the present study. 

In the following, we show two kinds of experiments. 
In the first experiment, we investigate the dependence of $\Delta$ on the film thickness by measuring the 
current-voltage ($I-V$) characteristic of the SET. For this experiment, smaller $E_J/E_c$ ratio 
is desirable for better accuracy in $\Delta$ (see below). 
In the second experiment, we investigate the gate modulation of the SET supercurrent with various thicknesses. 
Experimentally, it becomes easier to observe supercurrent, if the $E_J/E_c$ ratio of the SET is larger. 
Therefore we prepared two sets of SET devices: one with small $E_J/E_c$ ratio ($\sim 10^{-2}$) 
and the other with large $E_J/E_c$ ratio ($\sim 10^{-1}$), by changing the junction area and the oxidization time. 

\begin{figure}
\includegraphics[width=0.95\columnwidth,clip]{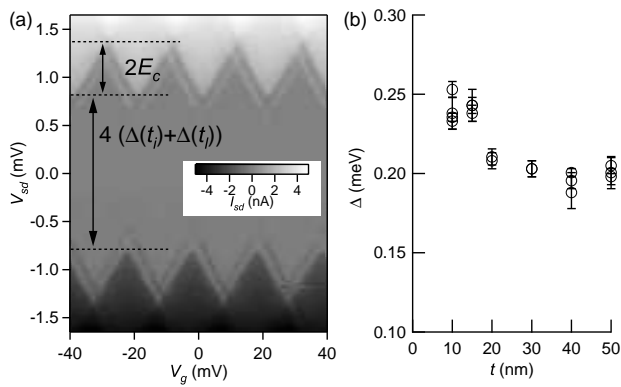}
\caption{\label{fig:fig2}
(a) Intensity plot of $I_{sd}$ as functions of $V_g$ and $V_{sd}$. 
The device has 40 nm thick island and 50 nm thick leads. 
(b) Observed superconducting energy gap of Al as a function of the film thickness. 
}
\end{figure}

In the first experiment, we use small $E_J/E_c$ ratio samples. 
They have typically normal source-drain resistance $R_{sd}=260$~k$\Omega$ and $E_c=340~\mu$eV. 
We fabricate SET devices with the thickness of the island ($t_i$) ranging from 10 to 50 nm. 
The thickness of the leads ($t_l$) is always 50 nm. Using a $^3$He refrigerator at temperatures of $\sim$ 300 mK, 
we measure the source-drain current ($I_{sd}$) as a function of the gate voltage ($V_g$) for different 
values of fixed source-drain bias voltage ($V_{sd}$), as it is exemplified in an intensity 
plot in Fig.~\ref{fig:fig2}a for the sample with $t_i=40$~nm. 
One can see a clear Coulomb blockade diamond structure: a sharp rise of the QP tunneling current 
above the Coulomb blockade threshold voltage $V_{th}$ which is modulated by $V_g$. 
From the modulation period, we estimate the capacitance between the gate electrode and the island $C_g$ to be 8.1 aF. 
The arrays of peaks and dips parallel to the diamond structure are due to Josephson-quasiparticle cycles.~\cite{Fulton89} 
As shown in the figure, $V_{th}$ is modulated between $2(\Delta(t_i)+\Delta(t_l))$ and 
$2(\Delta(t_i)+\Delta(t_l))+2E_c$, where $\Delta(t)$ is the superconducting gaps of the 
Al thin film with the thickness $t$. 
Thus, we can determine the value of $\Delta(t_i)$+$\Delta$(50 nm) 
from this measurement,~\cite{comment1} and $\Delta$(50 nm) can be determined from the similar measurement for 
the sample with $t_i$ of 50 nm, which turned out to be 0.2~meV on average. 
The derived $\Delta(t)$ is plotted in Fig.~\ref{fig:fig2}b. 
It is observed that $\Delta$ increases as $t$ is decreased, especially below 20 nm. 
This is qualitatively consistent with the previous reports.~\cite{Townsend72} 
The thinnest sample in this study gives 20 to 25{\%} higher $\Delta$ as compared to $\Delta$(50 nm). 
This variation of $\Delta$ is as high as the one achieved by oxygen doping in Ref.~\onlinecite{Aumentado04}, 
indicating that we can create a similar spatial profile of $\Delta$ in the SET by controlling the film thickness. 
In the next experiment, we check the effect of the $\Delta$ profile on the periodicity of the SET supercurrent. 

\begin{table}
\caption{\label{tab:parame}Sample parameters of large $E_J/E_c$ ratio samples.}
\begin{ruledtabular}
\begin{tabular}{ccccc}
Sample & $R_{sd}$ [k$\Omega$] & $E_c$ [$\mu$eV] & $E_J$ [$\mu$eV] & periodicity \\
\hline
10/50a & 52 & 145 & 27 & 2$e$ \\
10/50b & 53 & 141 & 27 & 2$e$ \\
10/30 & 38 & 139 & 37 & 2$e$ \\
40/50 & 32 & 114 & 40 & 2$e$ \\
50/10 & 47 & 160 & 30 & - \\
\end{tabular}
\end{ruledtabular}
\end{table}

\begin{figure}[hhh]
\includegraphics[width=0.95\columnwidth,clip]{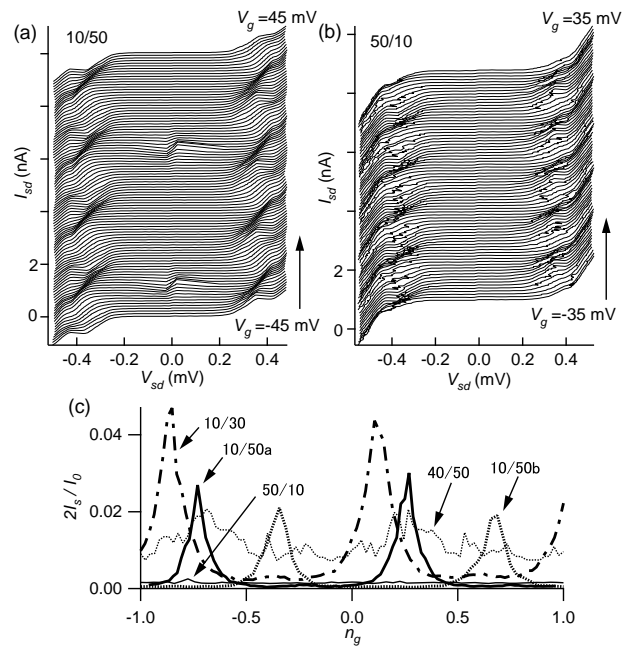}
\caption{\label{fig:fig3}
$I_{sd}-V_{sd}$ curves at different gate voltages for (a) the sample 10/50 and (b) the sample 50/10. 
Each curve is offset by 0.1 nA for clarity. 
(c) Gate charge dependence of the switching current for all the measured samples. 
Note that random offset charges have not been subtracted. 
}
\end{figure}

Based on the result of the first experiment, we fabricate SET devices (large $E_J/E_c$ ratio) 
with several combinations of $t_i$ and $t_l$, by which we will denote the samples as $t_i/t_l$ hereafter. 
Sample parameters are summarized in Table~\ref{tab:parame}. 
In all samples, the thinner layer was evaporated first. 
The measurements were performed in a dilution refrigerator at a base temperature of 50 mK 
using CuNi coaxial cables with $RC$ filters as dc lines. 
Using the current-biased configuration, we swept $I_{sd}$ at a rate of $\sim$0.1 nA/s 
and measured $V_{sd}$ for different values of fixed $V_g$. 
Here we show the results of such measurements for two samples, which have opposite $\Delta$ profiles: 
the sample 10/50a (higher $\Delta$ for the island) in Fig.~\ref{fig:fig3}a, 
and the sample 50/10 (higher $\Delta$ for the leads) in Fig.~\ref{fig:fig3}b. 
Each curve in the panels is offset by 0.1 nA for clarity. 
In Fig.~\ref{fig:fig3}a, current steps are observed near zero $V_{sd}$, to which we attribute the supercurrent. 
They are modulated by $V_g$ and the period is twice larger than that of QP tunneling current 
observed at higher $V_{sd}$, indicating that the supercurrent is 2$e$ periodic. 
In Fig.~\ref{fig:fig3}b, we observe the QP tunneling current modulated by $V_g$, 
but the supercurrent is hardly observed. 

From these measurements, we obtained the switching current $I_s$ and 
plotted as a function of the reduced gate charge $n_g=C_gV_g/2e$ for all the samples in Fig.~\ref{fig:fig3}c. 
Here, $I_s$ is normalized by the theoretical maximum $I_{0}/2$, where $I_{0}\equiv \pi\Delta_{eff}/2eR_N$ is the 
Ambegaokar-Baratoff critical current, $R_N$ is the normal-state resistance of the single junction, 
and $\Delta_{eff}=2\Delta(t_i)\Delta(t_l)/(\Delta(t_i)+\Delta(t_l))~$[\onlinecite{Baronebook}]. 
We assume that both junctions in the SET have same resistances. 
We observe clear 2$e$ periodicity in the samples 10/50a, 10/50b and 10/30. 
In all these samples, $\Delta$ of the island is about 20{\%} higher than that of the leads. 
In the sample 40/50, we still see the 2$e$ periodicity, 
but the contrast of the gate modulation becomes small. 
In the sample 50/10, the amplitude of the supercurrent peaks is very much suppressed and shows no clear modulation. 

The fact that we observe clear 2$e$ periodicity only in the samples where $\Delta$ of the island is higher 
than that of the leads is consistent with the previous report.~\cite{Aumentado04} 
Namely, in those samples, we create a spatial profile of $\Delta$ as shown in Fig.~\ref{fig:fig1}b. 
The energy gap difference $|\delta\Delta|$ ($\equiv |\Delta(t_l)-\Delta(t_i)|$) is about 40~$\mu$eV, 
which works as a barrier preventing nonequilibrium QPs created in the leads from entering the island. 
As discussed in Ref.~\onlinecite{Aumentado04}, observation of 2$e$ periodicity does not mean complete absence of QPs. 
In sample 10/50a, $\delta\Delta+\delta E_c^{01}$ becomes positive at $n_g>n_{g0}=0.32$, 
where $\delta E_c^{01}=4E_cn_g^2-4E_c(0.5-n_g)^2$ is the electrostatic energy difference 
between no QP and one QP states of the island.~\cite{comment3} 
It means that QP can tunnel into the box in the vicinity of the degeneracy point~(e.g. $0.32<n_g<0.68$). 
Although no clear sign of poisoning is seen, 
the small magnitude of 2$I_s/I_{0}$ may be the indication of this QP poisoning,~\cite{comment2}  
because our current ramping time is much longer than the inverse of the QP tunneling rate, 
which is reported to be $\sim10~\mu$s.~\cite{Aumentado04,Naaman06}
In sample 50/10, one QP state is favorable in a wider range, $0.13<n_g<0.87$, 
which leads to the poisoning of the supercurrent.
Although we expect the gate modulation of $I_s$ with 1$e$ periodicity, 
it is not clearly observed probably because the magnitude of $I_s$ is so small that 
it can be easily smeared out by the external noise. 

In summary, by measuring the $I-V$ characteristics of the SET, 
we have shown that $\Delta$ of Al depends on the film thickness. 
We have measured the gate modulation of the supercurrent of the SET devices, 
whose spatial profiles of $\Delta$ were controlled by the film thickness. 
Clear 2$e$ periodicity is observed only when the island is much thinner than the leads, 
demonstrating that the $\Delta$ profile controlled by the film thickness can help to suppress QP poisoning. 

The authors would like to thank T. Duty, D. B. Haviland, H. Im, A. O. Niskanen and M. Watanabe for fruitful discussions. 
This work was partially supported by the Japan Science
and Technology Corporation.


\begin{thebibliography}{21}
\bibitem[{com({\natexlab{a}})}]{comment1}
\bibinfo{note}{The data in Fig. 2b is obtained from the peak position of
  $dI_{SD}/dV_{SD}$.}
\bibitem[{com({\natexlab{b}})}]{comment3}
\bibinfo{note}{Strictly speaking, we also need to take into account $E_J$.
  However, with the present value of $E_J/E_c$ ratio, its effect on $n_{g0}$ is
  negligibly small ($<0.005$).}
\bibitem[{com({\natexlab{c}})}]{comment2}
\bibinfo{note}{Another reason for the observed small 2$I_s/I_{0}$ is
  insufficient filtering for the external noise in the measurement lines.}
\end{thebibliography}

\end{document}